\pdfoutput=1
\RequirePackage{ifpdf}
\documentclass[cits, twocolumn]{JINST}

\usepackage{amsmath}

\DeclareMathOperator{\he}{H}
\DeclareMathOperator{\nl}{Nl}

\title{Beam energy measurements for an experiment on~elastic $e^{\pm}p$~scattering at the VEPP--3 storage ring}

\author{V.V.~Kaminskiy$^{a,b}$\thanks{Corresponding author.}, %
A.V.~Gramolin$^a$, %
S.I.~Mishnev$^a$, %
N.Yu.~Muchnoi$^{a,c}$, %
V.V.~Neufeld$^a$, %
D.M.~Nikolenko$^a$, %
I.A.~Rachek$^a$, %
D.K.~Toporkov$^{a,c}$ %
and V.N.~Zhilich$^{a,c}$\\
\llap{$^a$}Budker Institute of Nuclear Physics, 630090 Novosibirsk, Russia\\
\llap{$^b$}Novosibirsk State Technical University, 630073 Novosibirsk, Russia\\
\llap{$^c$}Novosibirsk State University, 630090 Novosibirsk, Russia\\
E-mail: \email{V.V.Kaminskiy@inp.nsk.su}}

\abstract{A precise comparison of the cross sections for the elastic scattering of electrons and positrons on protons was performed recently at the VEPP--3 storage ring in Novosibirsk, Russia. To provide the proposed accuracy, the energies of the electron and positron beams used during the experiment needed to be carefully maintained. This paper describes the special beam energy measurement system used in this experiment. This system is based on measuring the maximum energy of laser photons backscattered from electron and positron beams. It provides continuous beam energy measurements with a total relative uncertainty of less than~$4.4 \cdot 10^{-5}$.}


\begin{document}

\section{Introduction}

The study of the electromagnetic form factors of the proton is crucial for the understanding of its internal structure. In addition, reliable knowledge of these form factors is essential for many areas of nuclear physics. However, a large discrepancy was recently observed between the ratio of the proton's electric to magnetic form factors measured with the polarization transfer method and the older results obtained using the Rosenbluth separation technique~\cite{PPNP.59.694}.

One of the most probable explanations of this discrepancy is the unaccounted contribution from hard two-photon exchange (TPE) to the elastic electron-proton scattering cross section~\cite{PPNP.66.782}. Until recently, it was assumed that the hard TPE effects were small, and they were usually neglected in the analysis of Rosenbluth measurements. However, the TPE contribution can be directly determined by measuring the ratio of positron-proton to electron-proton elastic scattering cross sections, $R = \sigma (e^+ p) / \sigma (e^- p)$. Currently, only old (from 1960's) experimental data on~$R$ exist, where the TPE contribution was found with low precision and for a limited kinematic coverage.

A new precise comparison of the $e^+ p$ and $e^- p$ elastic scattering cross sections was performed recently at the VEPP--3 storage ring in Novosibirsk, Russia~\cite{nucl-ex.0408020, NPBPS.225-227.216}. VEPP--3 can produce and store either electron or positron beams having energies from $0.35$ up to $2~\text{GeV}$ and moving in the same direction~\cite{VEPP}. The measurement of the ratio~$R$ at VEPP--3 has two phases: the first one performed in 2009 at a beam energy of $1.6~\text{GeV}$ and the second one (in 2011--2012) at a beam energy of $1.0~\text{GeV}$. In this experiment, the ratio~$R$ has been measured with the best achieved accuracy.

The experimental setup consisted of the VEPP--3 beam-line with an internal hydrogen gas target and a set of particle detectors to record the elastic scattering of electrons/positrons from the beam on the target protons. The typical data-taking run was the following: storing of the 30--50~mA beam (taking about 5 minutes for electrons and about 20 minutes for positrons); raising the beam energy from $0.35$ to~$1.0$ or~$1.6~\text{GeV}$; turning on the internal gas target; and data acquisition until the beam current drops below a value of about~$10~\text{mA}$. To reduce systematic errors, experimental conditions should be very similar for electron and positron runs. For this reason, the beam orbits, energies and average currents were kept nearly identical for $e^-$ and $e^+$ beams. Electron and positron runs were alternated regularly, and a special magnetization cycle was performed between them to suppress $e^-$ and $e^+$ beam energy inequality due to magnetic hysteresis of magnets. The typical working cycle consisted of one $e^-$ run and one $e^+$ run and took approximately 80~minutes in total, about half of this time spent on data acquisition.

The systematic uncertainty of~$R$ measured in the described experiment is estimated to be $3 \cdot 10^{-3}$ or better. A possible unaccounted difference between the electron and positron beam energies significantly contributes to this uncertainty. For example, a difference of $1~\text{MeV}$ leads to an uncertainty of about $2 \cdot 10^{-3}$. Therefore, it is necessary to continuously measure the beam energy and to keep it at the same level for different runs. The Compton backscattering of laser radiation appeared as the most convenient method for precise beam energy measurements in the discussed experiment. The development and experimental verification of this approach are described in~\cite{PRE.54.5657, NIMA.384.307, NIMA.384.293, JSR.5.392, NIMA.486.545, NIMA.598.23, ICFABDN.48.195, NIMA.659.21}. The method allows one to measure the electron/positron beam energy continuously and with a relative accuracy up to $2 \cdot 10^{-5}$~\cite{NIMA.659.21}.

\section{Concept of the method}

\begin{figure}
\centering
\includegraphics[width=0.5\linewidth]{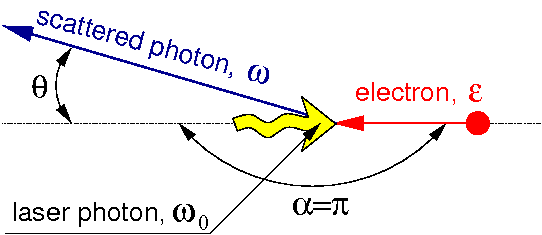}
\caption{The Compton backscattering process.}
\label{fig1}
\end{figure}

The beam energy measurement is based on the Compton scattering of low-energy laser photons on ultrarelativistic electrons/positrons~\cite{NIMA.598.23, ICFABDN.48.195, NIMA.659.21}. When a photon collides head-on with an electron/positron (such that $\alpha = \pi$, see figure~\ref{fig1}) and then is scattered on the angle~$\theta = 0$, its energy reaches the maximum allowable value,
\begin{equation}
\omega_{\text{max}} = \frac{4\gamma^2 \omega_0}{1 + 4\gamma \omega_0 / m_e} \approx 4\gamma^2 \omega_0.
\end{equation}
Having measured this energy~$\omega_{\text{max}}$, one can determine the beam energy~$\epsilon$ using the formula
\begin{equation}
\varepsilon = \frac{\omega_{\text{max}}}{2} \left(1 + \sqrt{1 + \frac{m_e^2}{\omega_0 \omega_{\text{max}}}} \, \right). \label{eq2.2}
\end{equation}

\begin{figure}
\centering
\includegraphics[width=0.7\textwidth]{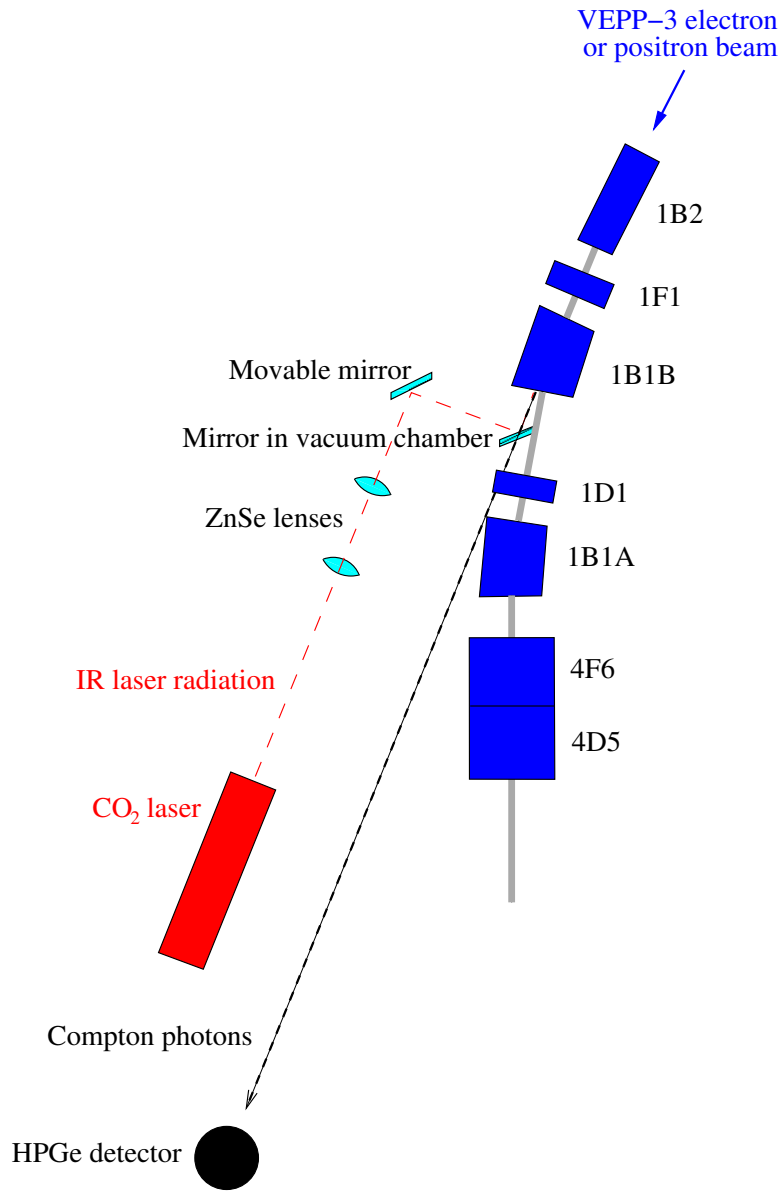}
\caption{Layout of the VEPP--3 beam energy measurement system.}
\label{fig3}
\end{figure}

The beam energy measurement procedure is the following. The monochromatic laser photons of energy $\omega_0 \approx 0.1~\text{eV}$ are injected into the vacuum chamber, where they interact with the electron/positron beam. The backscattered photons are detected using a high-purity germanium (HPGe) detector with an excellent energy resolution, $\Delta \omega / \omega \approx 10^{-3}$, and adequate detection efficiency for $\gamma$-quanta in the energy range of a few MeV. Simultaneously, the detector records radiation from several well-known monochromatic $\gamma$-quanta sources for the purpose of energy scale calibration. Then, a narrow edge corresponding to~$\omega_{\text{max}}$ is found in the acquired spectrum, the value of~$\omega_{\text{max}}$ is precisely measured, and the beam energy is calculated using the formula~(\ref{eq2.2}). Uncertainty in the obtained beam energy is determined mostly by the $\omega_{\text{max}}$ measurement error.

\section{Experimental setup}

The beam energy measurement system is placed near the VEPP--3 straight section opposite of the internal gas target section. The layout of the system is shown in figure~\ref{fig3}. $\text{CO}_2$ laser emission is injected into the VEPP--3 vacuum chamber using a special optical system. The interaction of photons with electrons/positrons occurs in the straight section between the 1B1B and 1B2 bending magnets. Compton-backscattered photons are detected by the HPGe detector.

\subsection{Laser and optical system}

The $\text{CO}_2$ laser GEM~Select~50 manufactured by Coherent,~Inc. is used as a source of monochromatic low-energy photons. It emits continuous (CW) radiation of variable power up to~$50~\text{W}$. The wavelength, $\lambda = 10.5910352~\mu\text{m}$, and the photon energy, $\omega_0 = 0.1170652~\text{eV}$, correspond to the 10P20 transition in the $\text{CO}_2$ molecule. The wavelength of the radiation is known with an accuracy of~$0.03~\text{ppm}$.

Special lenses are used to obtain maximum radiation density in the interaction region and, thus, to increase interaction efficiency. Two ZnSe lenses with the focus distance $f = 35~\text{cm}$ are placed at about $4~\text{m}$ from the interaction point with a gap between them of $60~\text{cm}$. The total path of the laser beam is approximately~$8~\text{m}$.

The laser system is placed below the median plane of the storage ring. The laser beam is reflected upward to the vacuum chamber by a special mirror. This mirror is installed with a movable support which allows one to adjust two angles of the mirror with a pair of remotely-controlled stepping motors.

The laser beam is injected into the vacuum chamber using a special laser-to-vacuum injection system. This has one flange connected to the VEPP--3 vacuum chamber for the laser beam, while another side is closed with a thin steel plate to transmit Compton-backscattered quanta. At the lower side is a transparent window made of ZnSe. Inside the injection system, a thin water-cooled adjustable copper mirror is installed which reflects the laser beam towards the interaction region and transmits the Compton quanta. The entire injection system is maintained at high vacuum.

\subsection{HPGe detector}

The detector GC2518 manufactured by Canberra is used for the beam energy measurement system. It is a coaxial closed-ended HPGe detector of $120~\text{cm}^3$ volume. It is cooled by liquid nitrogen in a 30 liter dewar. Its standard energy resolution for the $1.33~\text{MeV}$ line of~$^{60}\text{Co}$ is $1.8~\text{keV}$ (FWHM). The detector is connected to the multi-channel analyzer (MCA) ORTEC DSpec~Pro, which produces pulse height spectra and transfers data to the computer through a USB port.

Since the detector is located in the accelerator hall, the background radiation from the beam (bremsstrahlung from residual gas and electromagnetic showers caused by beam losses) and a positron injector (electromagnetic showers and neutrons) during runs is very high. For this reason, the detector is shielded by 5~cm thick lead blocks and 5~cm thick polyethylene blocks. At the side nearest the injector, the detector is additionally protected by one layer of lead and one layer of polyethylene blocks. The opposite to the injector side and the entrance window for $\gamma$-quanta from Compton backscattering and from calibration isotopes are covered by only one layer of polyethylene blocks.

Gamma-active isotopes are placed near the detector end-cap. Also, a precision pulse generator (model PB-5 from Berkeley Nucleonics~Corp.) is used for calibration of the detector energy scale. It is connected to a special input of the HPGe detector. The generator pulse amplitude is controlled remotely by PC.

\section{Data acquisition system}

The spectra acquisition procedure is the following. The HPGe detector spectra are read every few seconds, and counting rates are calculated. If the requested acquisition time has elapsed, the number of events has reached the required value, or the data-taking run has started or ended, the current spectrum is saved to a file and the next spectrum acquisition cycle is launched. Simultaneously, VEPP--3 parameters (beam polarity, beam current, and nominal beam energy) and the detector counting rates are obtained and written to a file for further analysis.

\begin{figure}
\centering
\includegraphics[width=0.7\textwidth]{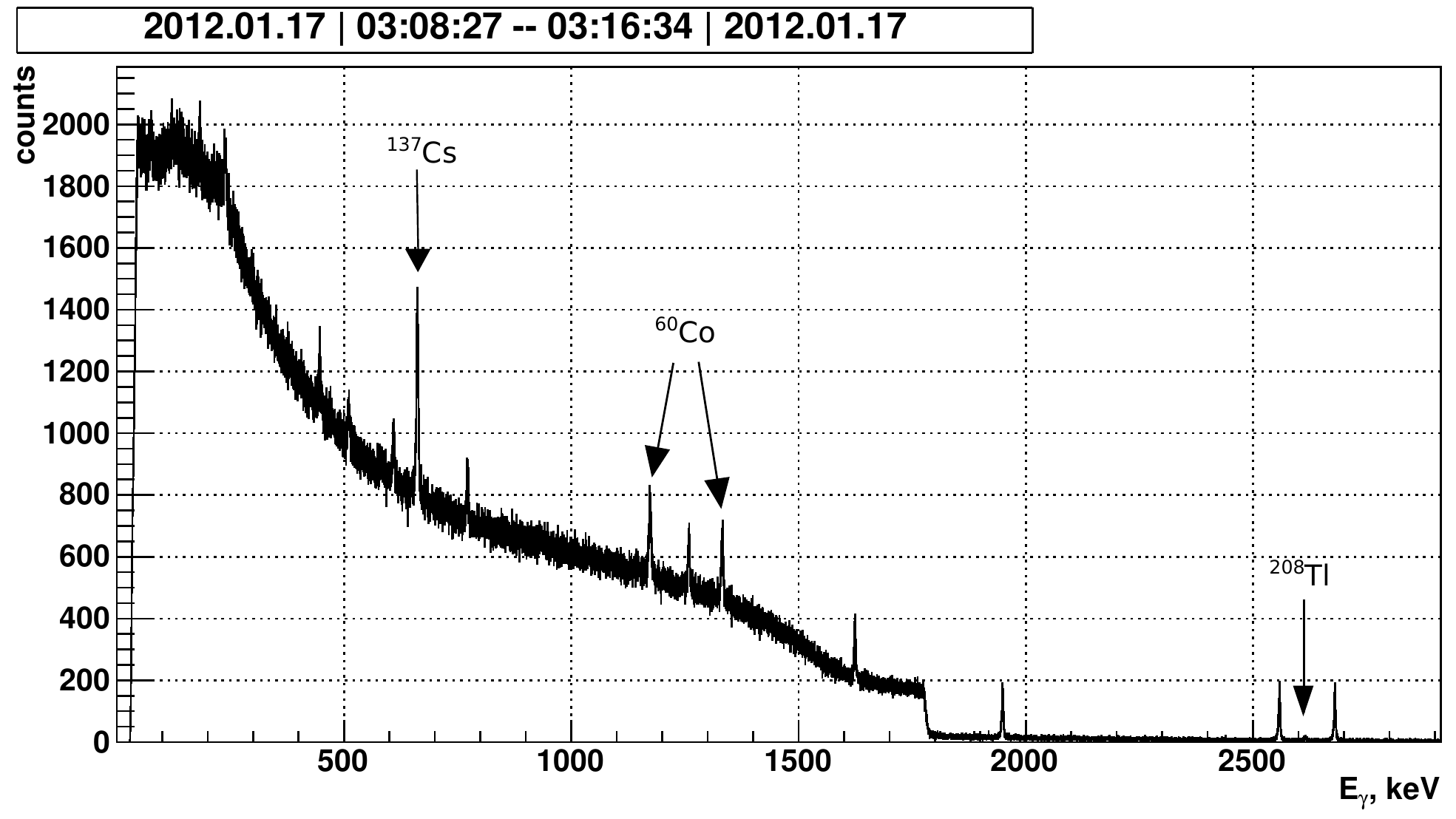}
\caption{A typical energy spectrum obtained by the HPGe detector. The peaks from the standard isotopes and from the precision pulse generator are seen, as well as the Compton edge (at $E_{\gamma} = 1780~\text{keV}$).}
\label{fig2}
\end{figure}

When the current spectrum acquisition cycle has finished, another program processes the spectrum: calibrates the energy scale, finds the Compton edge (if it presents), and calculates the beam energy. Then, the beam energy is written into a database. A typical energy spectrum obtained by the HPGe detector is shown in figure~\ref{fig2}.

During data acquisition, the movable mirror is adjusted automatically to provide a maximum overlap of the laser beam and the electron/positron beam, using the detector counting rate as feedback.

More details about a similar data acquisition system can be found in~\cite{NIMA.659.21}.

\section{Data processing}

The aim of the spectrum-processing procedure is to measure~$\omega_{\text{max}}$ and calculate the beam energy. To determine $\omega_{\text{max}}$ precisely, one needs to know the response function of the detector and to properly calibrate its energy scale.

Monochromatic $\gamma$-quanta always give a non-monochromatic response at the detector. To gain maximum accuracy in the determination of $\gamma$-radiation energy in the case of the HPGe detector having excellent resolution, the asymmetry of the response should also be considered. The direct way to determine the response function is to observe the full-energy peak from the monochromatic $\gamma$-quanta emitted by a known radioactive isotope. The following parametrization was used for the response function:
\begin{equation}
f (\omega, E_{\gamma}, \sigma, \xi) = \frac{N}{\sigma \sqrt{2\pi}} \cdot \left \{ 
\begin{array}{ll}
\exp {\left(-\frac{(\omega - E_{\gamma})^2}{2\sigma^2}\right)},                & \omega > E_{\gamma} - \sigma \xi, \\
\exp {\left(\frac{\xi^2}{2} + \frac{\xi(\omega - E_{\gamma})}{\sigma}\right)}, & \omega < E_{\gamma} - \sigma \xi, \\
\end{array}
\right. \label{eq5.1}
\end{equation}
where~$N$ is a normalizing factor, $E_{\gamma}$~is the energy of the peak centroid, $\sigma = \text{FWHM}/2.36$ is the total detector resolution, and $\xi$~is an asymmetry parameter.

Calibrating the energy scale means converting channels of the detector multi-channel analyzer to the initial photon energy in units of keV. In the procedure used, the calibration is comprised of two stages. First, a linear scale is assumed and linear coefficients for conversion are found. Then, the energy of interest is corrected using a nonlinearity function obtained with a precision pulse generator.

\begin{figure}
\centering
\includegraphics[width=0.7\textwidth]{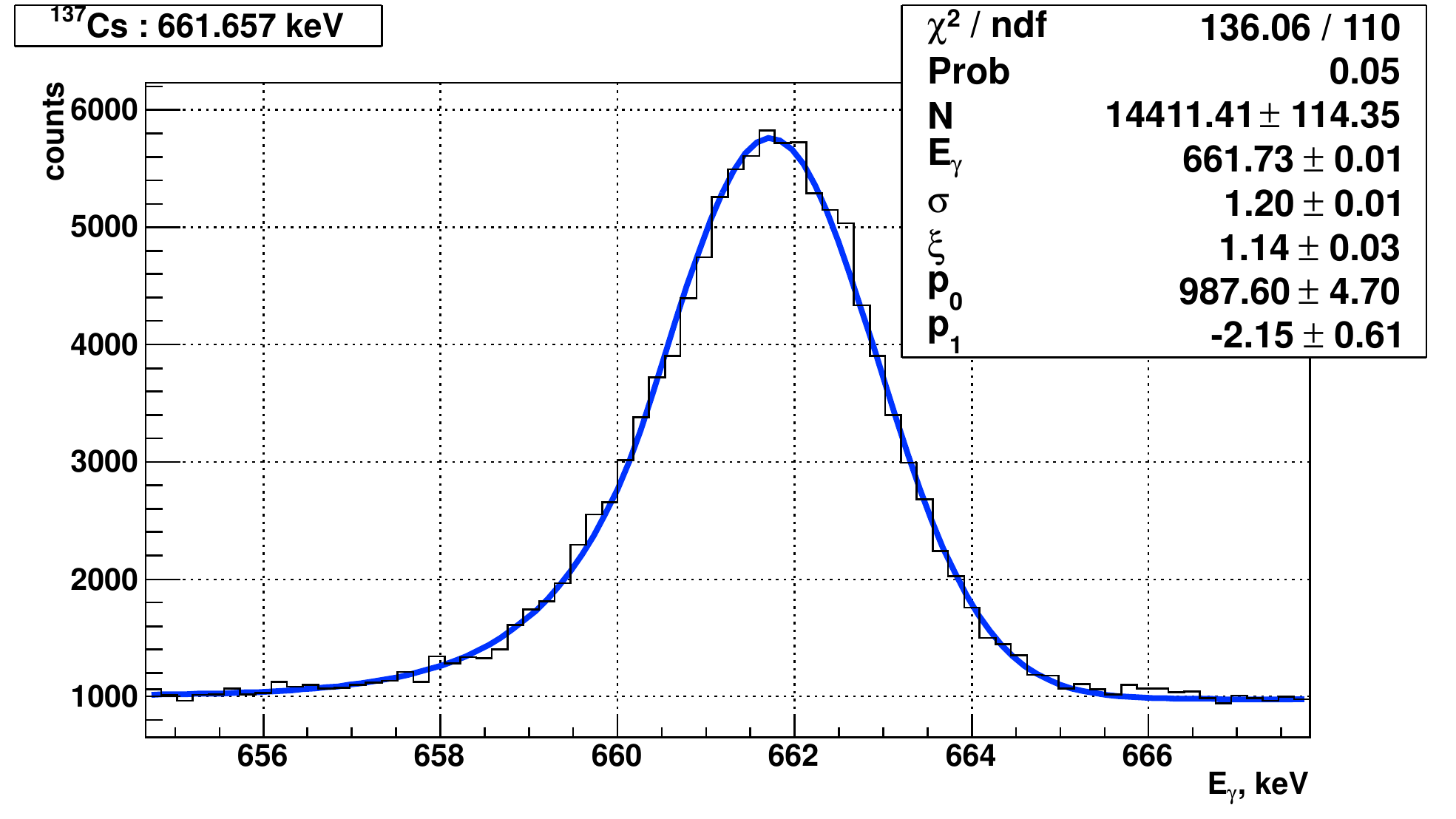}
\includegraphics[width=0.7\textwidth]{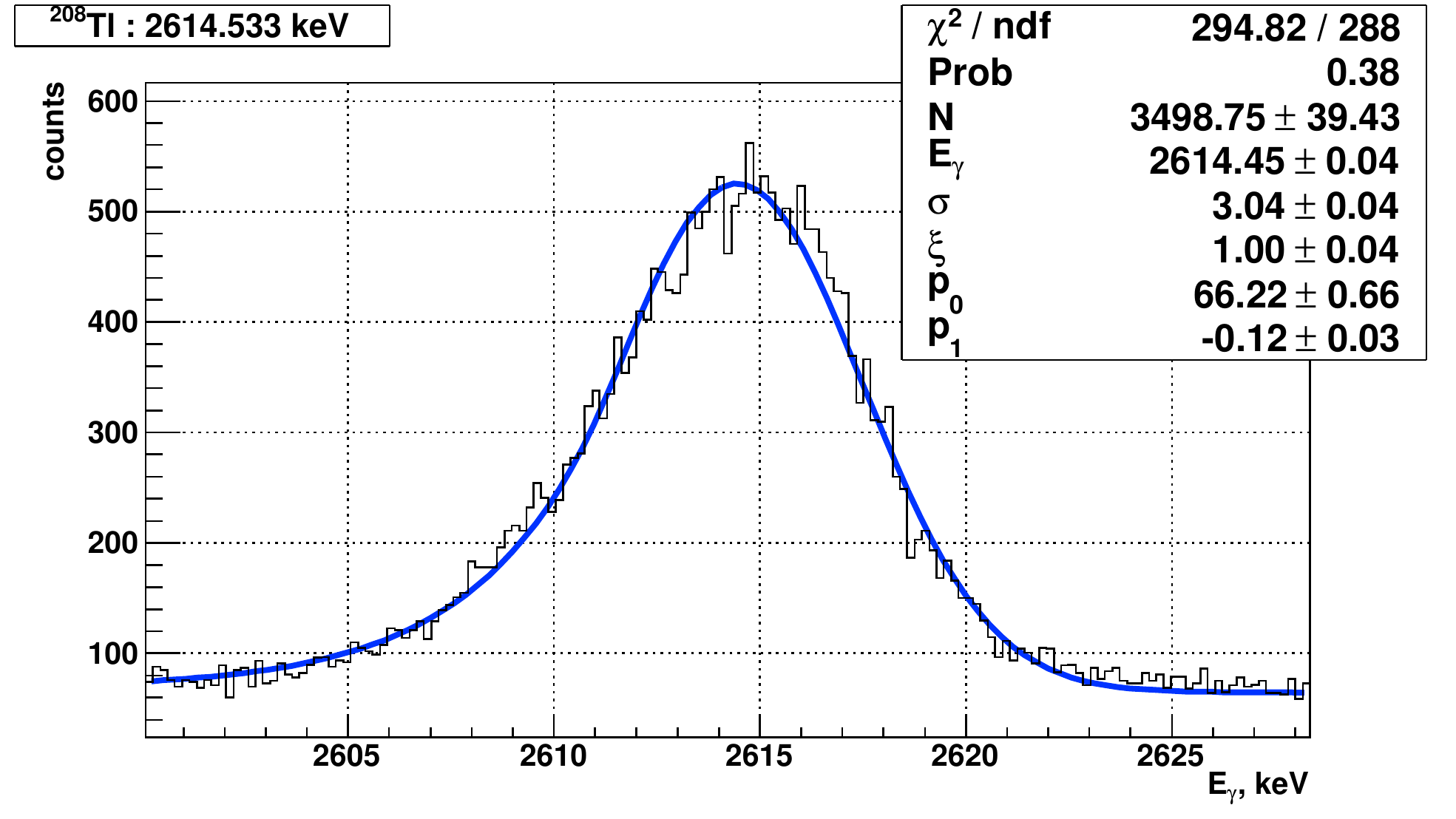}
\caption{Some peaks used for calibration, fitted with the response function.}
\label{fig4}
\end{figure}

The obvious way to perform an absolute calibration of the detector scale is to use $\gamma$-lines with precisely measured energies from the standard isotopes. The following well-known intensive $\gamma$-lines~\cite{NIMA.422.525} were used in the described experiment (shapes of some peaks are shown in figure~\ref{fig4}):
\begin{align*}
^{137}\text{Cs}, \quad E_{\gamma} &= \phantom{0}661.657 \pm 0.003~\text{keV}; \\
^{60}\text{Co}, \quad E_{\gamma} &= 1173.228 \pm 0.003~\text{keV}; \\
^{60}\text{Co}, \quad E_{\gamma} &= 1332.492 \pm 0.004~\text{keV}; \\
^{208}\text{Tl}, \quad E_{\gamma} &= 2614.533 \pm 0.013~\text{keV}.
\end{align*}

In detail, the calibration and data-processing procedure consists of the following 7~steps.
\begin{enumerate}
\item The spectra processing is performed using the ROOT software library. The $x$-axis of the spectrum is converted from MCA channels $N_{\text{MCA}}$ to energy~$\omega$ expressed in keV using predefined linear conversion coefficients: $\omega = Z + G \cdot N_{\text{MCA}}$.

\item All of the peaks in the spectrum are found and then identified using a list of $\gamma$-lines of interest. The identified peaks are least-squares fitted with the sum of the response function~(\ref{eq5.1}) and the linear background function:
\begin{equation}
f(\omega, E_{\gamma}, \sigma, \xi) + p_0 + p_1 \omega,
\end{equation}
where $E_{\gamma}$, $\sigma$, $\xi$, $p_0$, and $p_1$ are free fit parameters.

\item \label{item_sigma_xi} The dependencies of $\sigma$ and $\xi$ versus energy are plotted and fitted (see figures~\ref{fig5} and~\ref{fig6}). The graph of $\sigma$ versus $E_{\gamma}$ is fitted with the function 
\begin{equation}
\sigma(E_{\gamma}) = \sqrt{K_0^2 + F \epsilon_{\text{pair}} E_{\gamma}},
\end{equation}
where~$K_0$ corresponds to the noise contribution, from~$0.7$ to $1.4~\text{keV}$, $F$ roughly corresponds to the Fano factor, from $0.12$ to $1.0$, and $\epsilon_{\text{pair}}$ is the average energy required to create an electron-hole pair, $2.96 \cdot 10^{-3}~\text{keV}$. $K_0$ and~$F$ are free fit parameters. The energy dependence of the asymmetry parameter~$\xi$ is approximated by an empirical function $p_0 + p_1 \cdot \exp{(-p_2 \omega)}$. Thus, the response function is known in the available energy range.

\begin{figure}
\centering
\includegraphics[width=0.7\textwidth]{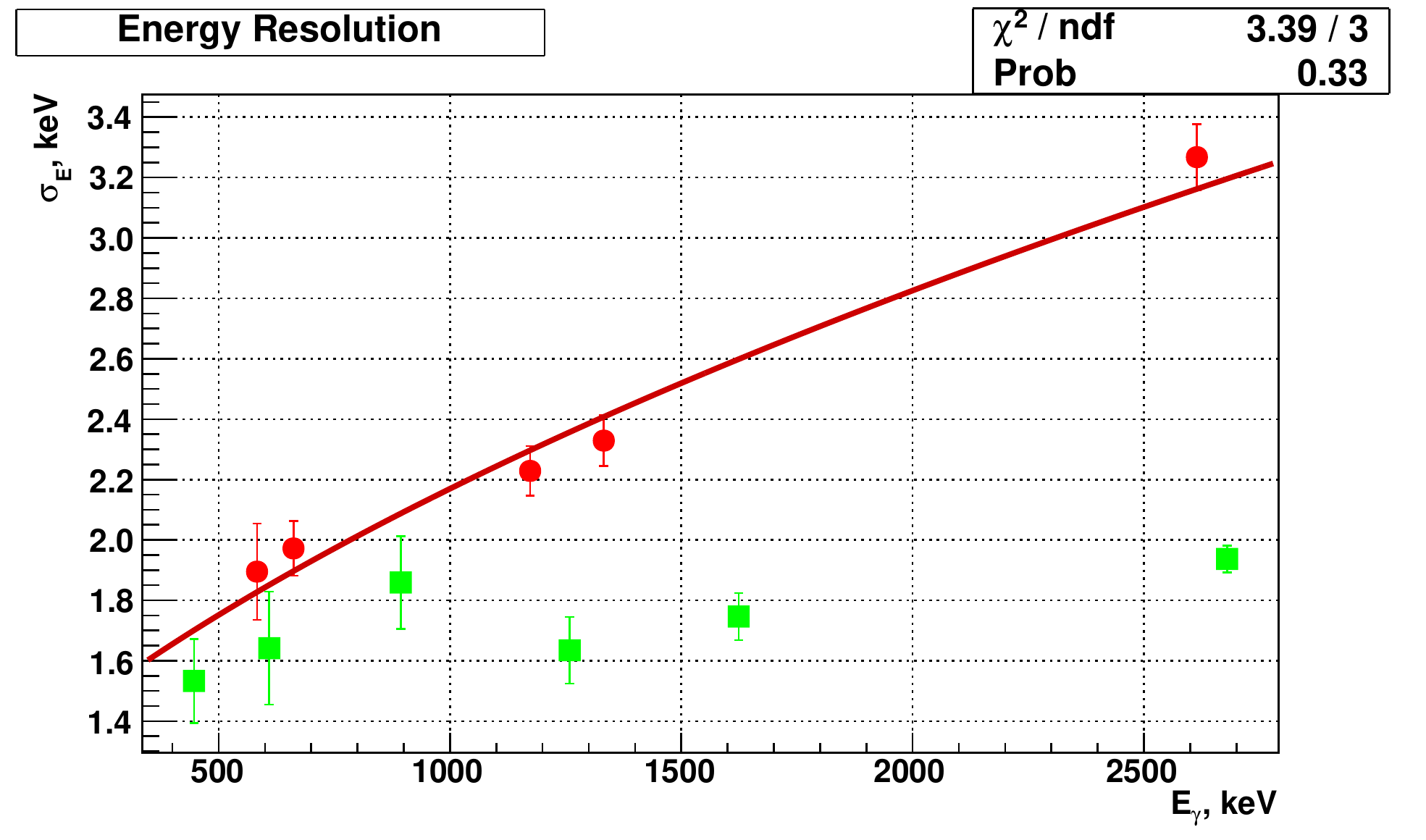}
\caption{Energy resolution~$\sigma$ of the HPGe detector. The red circles represent peaks from the isotopes and the green squares those from the precision pulse generator. The graph for isotope peaks is fitted with the function $\sigma(E_{\gamma}) = \sqrt{K_0^2 + F \epsilon_{\text{pair}} E_{\gamma}}$.}
\label{fig5}
\end{figure}

\begin{figure}
\centering
\includegraphics[width=0.7\textwidth]{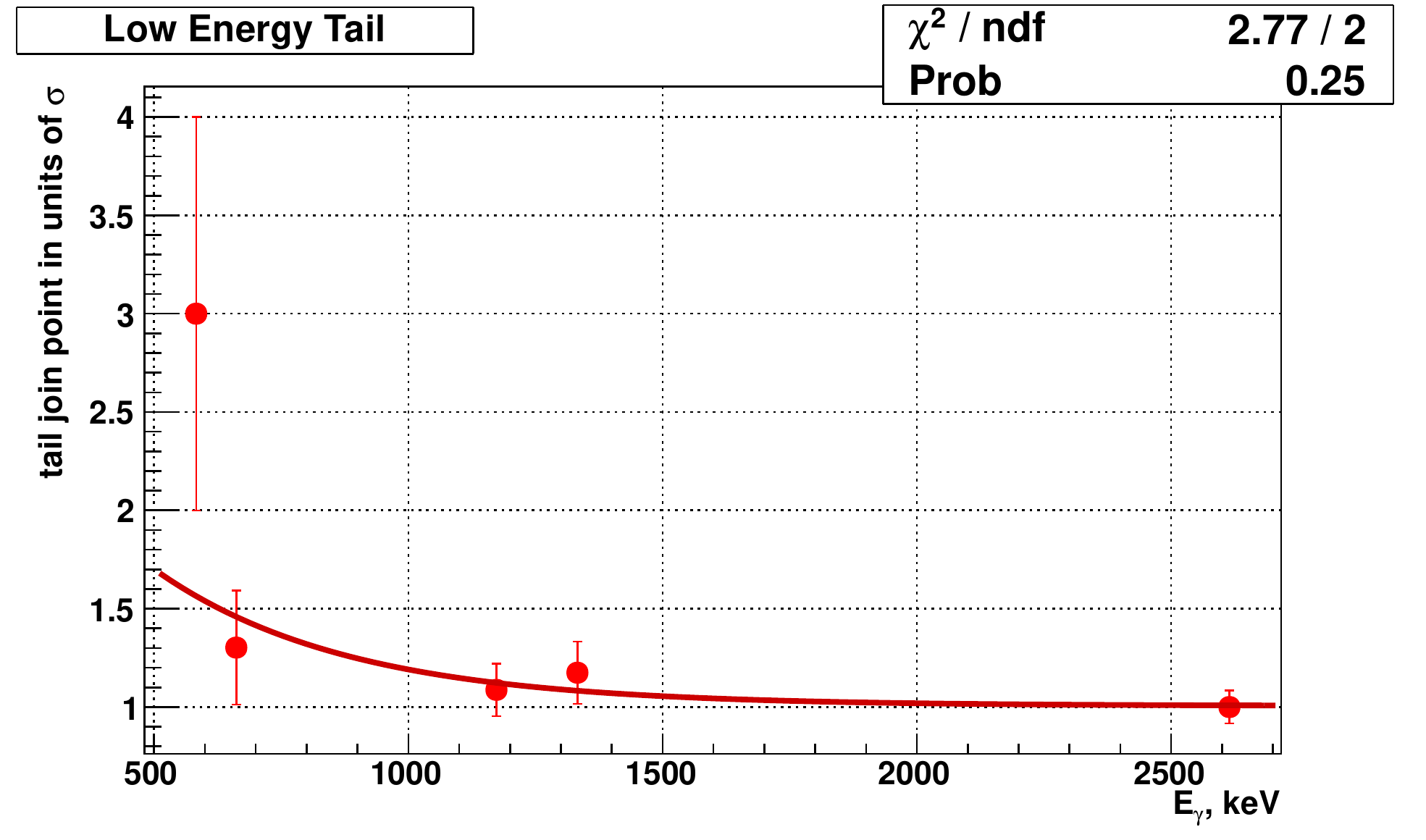}
\caption{Asymmetry parameter~$\xi$ as a function of energy~$E_{\gamma}$.}
\label{fig6}
\end{figure}

\begin{figure}[h]
\centering
\includegraphics[width=0.7\textwidth]{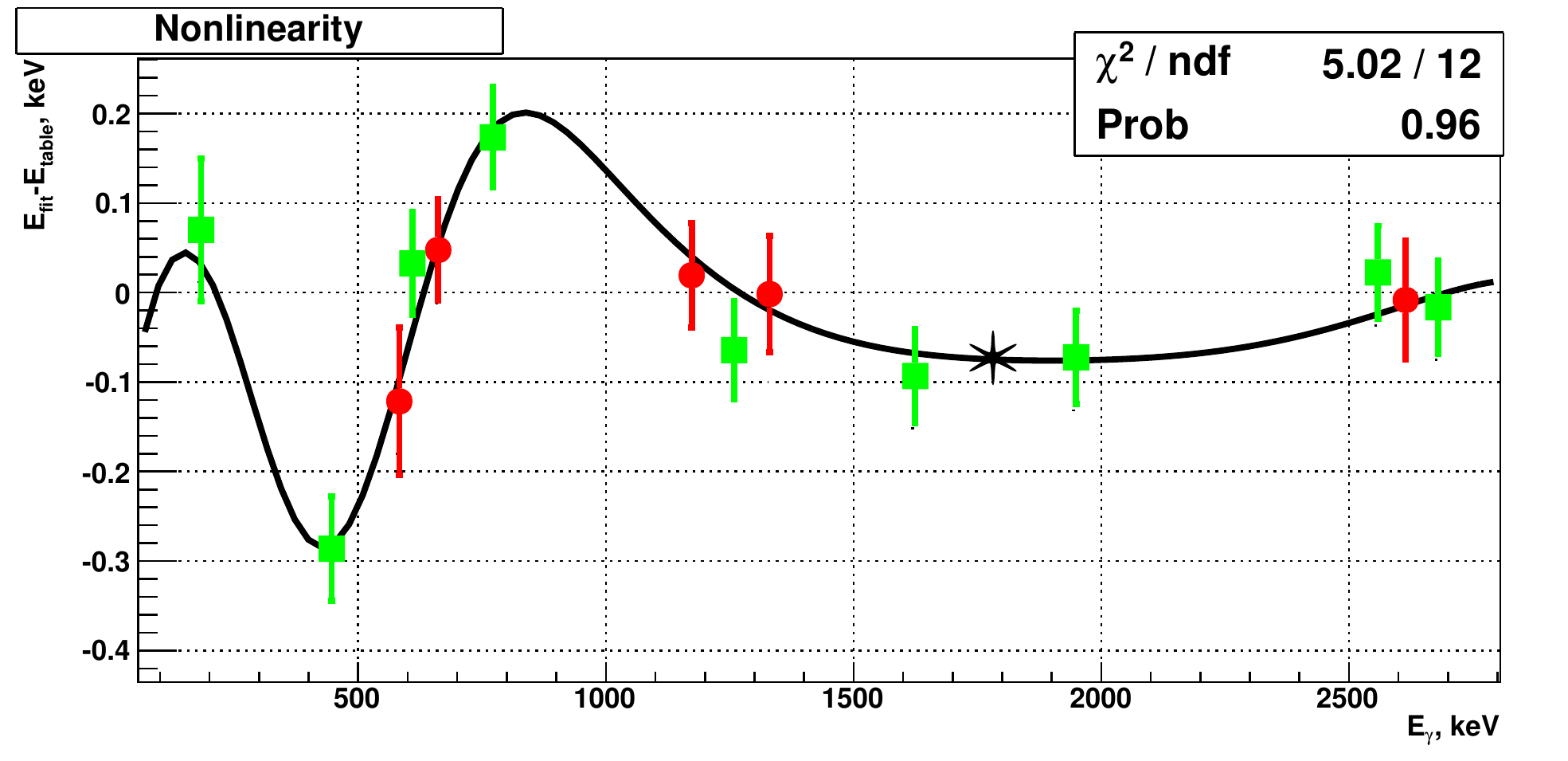}
\caption{Nonlinearity of the HPGe detector. The red circles represent peaks from the isotopes and the green squares represent one from the precision pulse generator. The asterisk represents the Compton edge energy. The thin black curve is a smooth nonlinearity function obtained with the precision pulse generator.}
\label{fig7}
\end{figure}

\item The difference between peak positions and their reference energies is plotted (see figure~\ref{fig7}). This difference can be treated as the nonlinearity. After fitting this plot with a linear function, the more precise linear $N_{\text{MCA}} \to \omega$ conversion coefficients are found. The same procedure is performed for the peaks of the precision pulse generator, using an equivalent energy: $\omega_{\text{gen}} = p_0 + p_1 U$, where $U$ is the generator voltage. 

\item When the nonlinearity graph of $\gamma$-lines from isotopes is obtained, it is fitted by the nonlinearity function $G \cdot \nl [(\omega - Z)/G]$, where the linear coefficients are free fit parameters and $\nl (N_{\text{MCA}})$ is a ninth-degree polynomial representing a direct MCA nonlinearity measurement with the precision pulse generator. A spectrum with 80 generator voltages over the whole MCA range is acquired and processed according to the algorithm similar to those of $\gamma$-lines. Figure~\ref{fig7} shows that the residual nonlinearity for isotopes and for the generator have the similar shapes, meaning that the nonlinearity is mostly caused by~MCA.

Now, the linear scale is found using isotopes and the nonlinear correction can be made with the help of the precision pulse generator.

\begin{figure}
\centering
\includegraphics[width=0.7\textwidth]{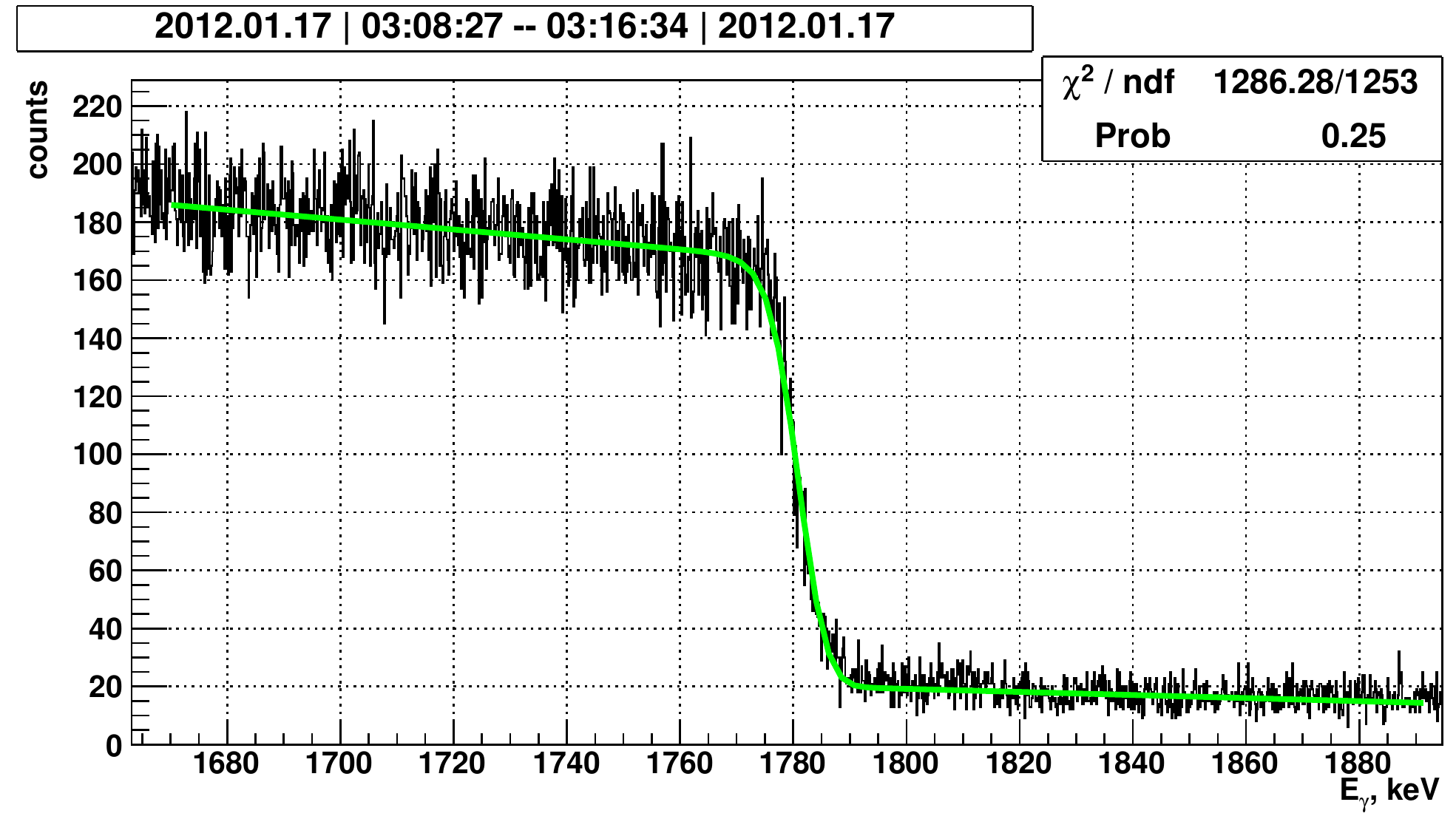}
\caption{The Compton edge in the HPGe detector spectrum. It is fitted by a convolution of a step-like function with the response function.}
\label{fig8}
\end{figure}

\item Finally, the Compton edge is fitted by the function
\begin{gather}
S (\omega, \omega_{\text{max}}, A, \sigma_{s}, \sigma, \xi, a_{\text{bg}}, b_{\text{bg}}) \nonumber \\
{} = A \iint\limits_{-\infty}^{+\infty} \he \left(\omega_{\text{max}} - \omega'\right) \, \exp{\left(-\frac{(\omega' - \omega)^2}{2\sigma_{s}^{2}}\right) d\omega' \: f(\omega'', \omega', \sigma, \xi) \, d\omega'' + a_{\text{bg}}*\omega + b_{\text{bg}}},
\end{gather}
where~$f$ is the response function~(\ref{eq5.1}), $\he \left(\omega_{max} - \omega'\right)$ is the Heaviside step function ($\he = 1$ if $\omega' < \omega_{\text{max}}$, otherwise 0), $\sigma_{s}$ is widening of $\omega_{\text{max}}$ due to the beam energy spread, and~$A$ is a normalizing factor. Literally, $S$ is a convolution of Compton spectrum from a strictly monochromatic electron beam (expressed as the Heaviside function) with a Gaussian due to the beam energy spread, with the detector response function. The Compton spectrum edge fitted with this function is shown in figure~\ref{fig8}. The parameters $\sigma$ and $\xi$ are extracted from graphs described in item~\ref{item_sigma_xi} of the procedure. The Compton edge energy is known from the fit parameters and then is corrected using the nonlinearity function. The beam energy is then calculated using equation~(\ref{eq2.2}). The energy is then corrected by +45~keV at 1.6~GeV and by +6~keV to take into account energy loss due to synchrotron radiation between the point of energy measurement and the hydrogen target.

\item The total uncertainty of the beam energy for each run is estimated by
\begin{gather}
\Delta \epsilon_{\text{tot}} = \frac{\varepsilon}{2} \frac{\Delta \omega_{\text{max}}}{\omega_{\text{max}}} \nonumber \\
{} = \frac{\varepsilon}{2\omega_{\text{max}}} \sqrt{(\Delta \omega_{\text{max}})_{\text{stat}}^2 + \left(\frac{\partial \nl (\omega_{\text{max}})}{\partial \omega} \cdot (\Delta \omega_{\text{max}})_{\text{stat}}\right)^2 + (\Delta G \cdot \omega_{\text{max}})^2 + (\Delta Z)^2},
\end{gather}
where $\Delta \omega_{\text{max}}$ is a worst-case estimate of Compton edge energy uncertainty, $(\Delta \omega_{\text{max}})_{\text{stat}}$ is the statistical uncertainty of the Compton edge energy, and the terms with $\partial \nl (\omega_{\text{max}}) / \partial \omega$, $\Delta G$, $\Delta Z$ represent uncertainties caused by the nonlinearity function. $(\Delta \omega_{\text{max}})_{\text{stat}}$ is defined from the edge fitting, $\Delta G$, $\Delta Z$ are defined from the residual nonlinearity fitting.
\end{enumerate}

\section{Results of the measurements}

During the first phase of the experiment (at the beam energy of $1.6~\text{GeV}$), a typical accuracy obtained in a 15--20 minute run was $70~\text{keV}$, providing a relative energy resolution of $\Delta E/E = 4.4 \cdot 10^{-5}$. The results for the second phase of the experiment (at the beam energy of $1.0~\text{GeV}$) are $40~\text{keV}$ and $4.0\cdot 10^{-5}$ respectively. The improvement in the second phase is achieved by the introduction of the precision pulse generator. The accuracy obtained substantially exceeds the minimal requirements for the $e^+/e^-$ beams' energy difference in the TPE experiment at \mbox{VEPP--3} ($\Delta E < 1~\text{MeV}$). Therefore, the measurements described not only ensure the beams' energy difference to be inside a permissible window, but allow us to further suppress the systematic errors, related to the time-variation of beam energy, by applying energy-dependent corrections during the off-line analysis of the experimental data.

It should be emphasised that this accuracy is reached within 20-minutes run while in other experiments described in~\cite{PRE.54.5657, NIMA.384.307, NIMA.384.293, JSR.5.392, NIMA.486.545, NIMA.598.23, ICFABDN.48.195, NIMA.659.21} the better accuracy reached within 1 or 2 hours run. Moreover, in most of them the accuracy was approved with another precise energy measurement method (resonant depolarisation, J/$\psi$ scan).

\begin{figure}
\centering
\includegraphics[width=1\textwidth]{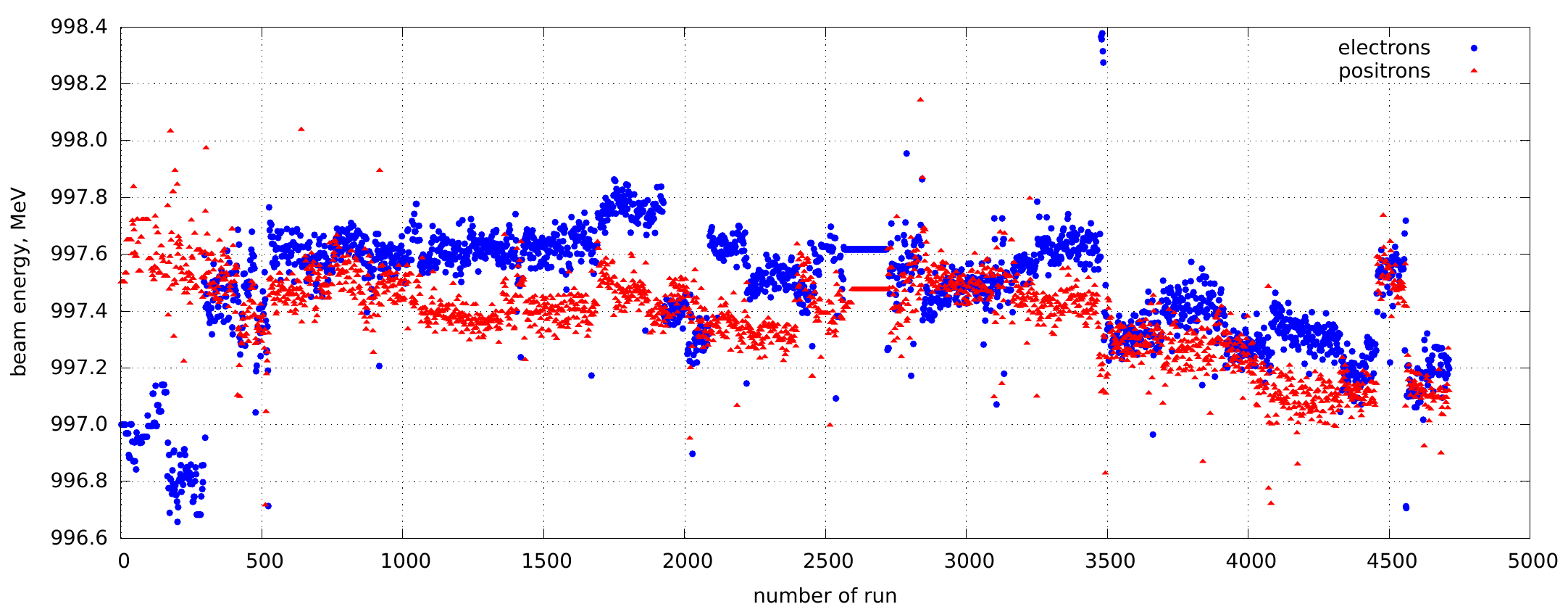}
\includegraphics[width=1\textwidth]{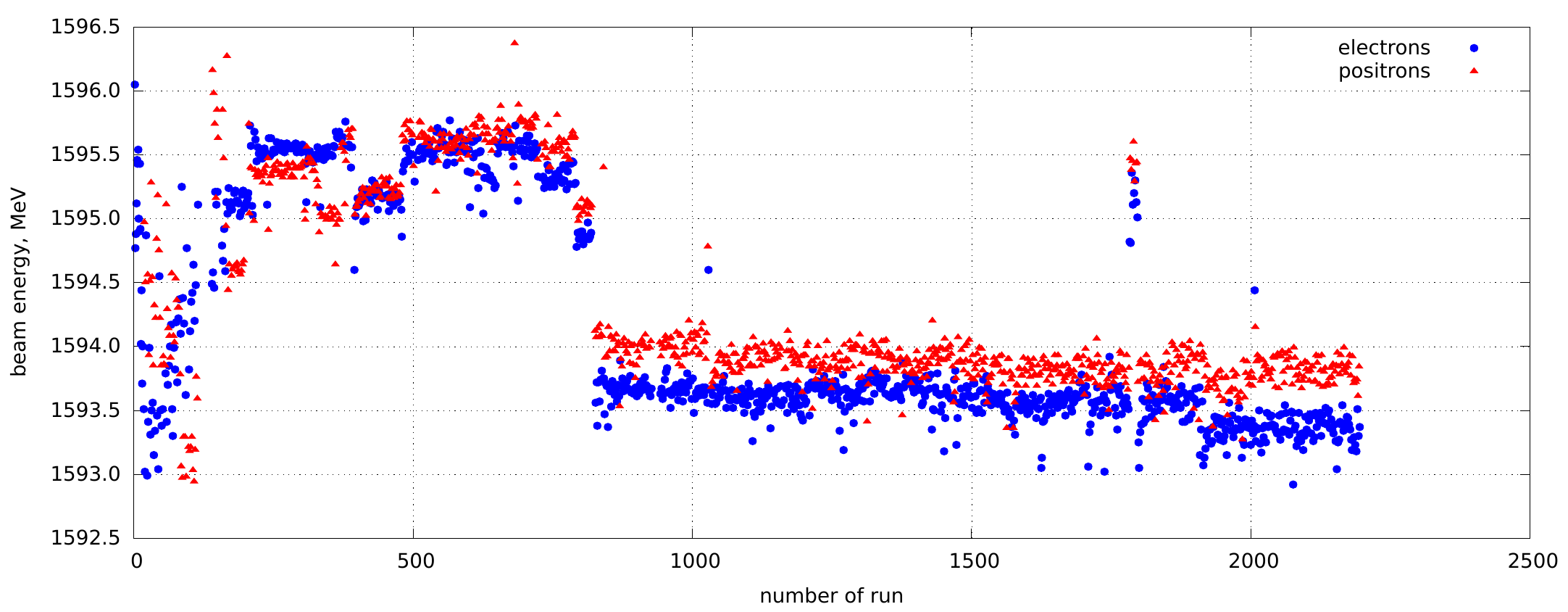}
\caption{Measured beam energy versus run number (with electron or positron beam): top panel for the first phase of the experiment, bottom panel for the second one. The blue circles correspond to electron beams and the red triangles to positron beams.}
\label{fig9}
\end{figure}

\begin{figure}
\centering
\includegraphics[width=1\textwidth]{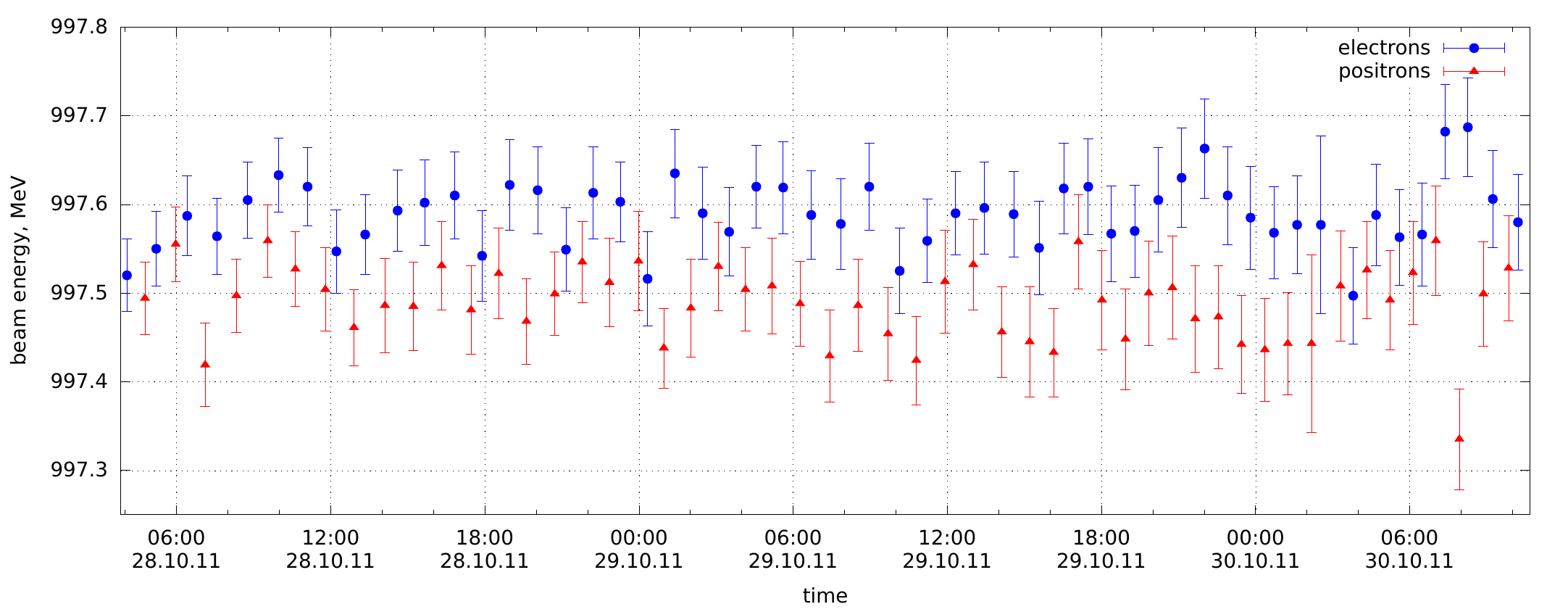}
\caption{Short-term behaviour of the beam energy for the first phase of the experiment. The blue circles correspond to electron beams, and the red triangles to positron beams.}
\label{fig10}
\end{figure}

The long-time behaviour of the measured beam energy during the two phases of the experiment is shown in figure~\ref{fig9}. As can be seen, each experimental phase starts with a period of tuning of the VEPP--3 regimes, characterized by a noticeable ``drifting'' and ``jumping'' of the beam's energy, followed by a long time of nearly stable conditions. Beam energy measurements by Compton backscattering were very useful at the stage of the adjustment of the VEPP--3 operating regimes. 
One can see that the difference between positron and electron energies does not exceed $400~\text{keV}$ for most of the time. 
An example of the short-term behaviour of the measured beam energy during a period of stable operation is demonstrated in figure~\ref{fig10}.

It seems that magnetic hysteresis was not suppressed completely. So, field integral for electrons and positrons was different, which caused different energies. Also different $e^-$ and $e^+$ beam positions at the point of energy measurement were observed (but was not studied properly), using the laser beam as ``a probe''. It could occur due to different lengths of electron and positron orbits, and the same beam positions near the target. As the energy measurement has high enough accuracy and it was performed for each run and the beam positions near the target was controlled with high enough accuracy, no thorough study of the systematic $e^-$ and $e^+$ beams difference was needed.

\section{Conclusion}

The VEPP--3 beam energy measurement system based on Compton backscattering was assembled and operated in an experiment on elastic $e^{\pm} p$~scattering. The relative accuracy achieved in each of the data-taking runs is better than $4.4 \cdot 10^{-5}$. Thus, the described system made it possible to control the energies of $e^+$ and $e^-$ beams with high accuracy and to perform at VEPP--3 the precise measurement of the cross section ratio $R = \sigma (e^+ p) / \sigma (e^- p)$.

\acknowledgments

The authors are grateful to the staff of VEPP--3 for the excellent performance of the storage ring during the experiment. The work was supported by the Ministry of Education and Science of the Russian Federation.

\end{document}